# A Complexity View of Markets with Social Influence

Xi Chen*
Department of Computer Science
Columbia University

Shang-Hua Teng[†]
Computer Science Department
University of Southern California.

**Abstract**

In this paper, inspired by the work of Megiddo on the formation of preferences and strategic analysis, we consider an early market model studied in the field of economic theory, in which each trader's utility may be influenced by the bundles of goods obtained by her social neighbors. The goal of this paper is to understand and characterize the impact of social influence on the complexity of computing and approximating market equilibria.

We present complexity-theoretic and algorithmic results for approximating market equilibria in this model with focus on two concrete influence models based on the traditional linear utility functions. Recall that an Arrow-Debreu market equilibrium in a conventional exchange market with linear utility functions can be computed in polynomial time by convex programming. Our complexity results show that even a bounded-degree, planar influence network can significantly increase the difficulty of equilibrium computation even in markets with only a constant number of goods. Our algorithmic results suggest that finding an approximate equilibrium in markets with hierarchical influence networks might be easier than that in markets with arbitrary neighborhood structures. By demonstrating a simple market with a constant number of goods and a bounded-degree, planar influence graph whose equilibrium is PPAD-hard to approximate, we also provide a counterexample to a common belief, which we refer to as the myth of a constant number of goods, that equilibria in markets with a constant number of goods are easy to compute or easy to approximate.

*Most of the work was done while the author was a postdoc at University of Southern California, where this research was supported by a USC startup fund.

[†]This research is supported by an NSF grant CCF-0964481.

# 1  Introduction

In mathematical economics, the general equilibrium theory has laid the foundation for competitive pricing [1, 24]. This theory was based on the supply-equal-demand principle of Adam Smith and Léon Walras [26]. For pricing in an exchange economy, the demand of a trader is usually modeled by a utility function which assigns each bundle of goods a non-negative value. An equilibrium price then leads the system to an efficient allocation of goods among traders. The pioneering equilibrium theorem of Arrow and Debreu [1] asserts the existence of market equilibria for a very general model of exchange markets.

In the traditional exchange model [1, 24], each trader's utility depends only on the bundle of goods she obtains after the exchange. However, this view of demand may have some limitations — many people's interests and value may be influenced by their social interactions [20]. For example, the more friends with iPhones that one has, the cheaper potentially it is for one to talk to them using an iPhone (due to the in-network service), and hence having an iPhone may be more valuable. In the age of ubiquitous social networks, it might be desirable to model markets and exchanges where traders' valuation and utilities are influenced by what their social neighbors have.

In this paper, we consider a market model in which each trader's utility is potentially influenced by the bundles of goods possessed by her social neighbors. The mathematical properties of this market model and its extensions have been extensively studied in economic theory, for example by Föllmer [14], Evstigneev and Takasar [13, 12], and Horst and Scheinkman [16]. Formally, a market in this model is defined by traders' initial endowments of goods, a social network among traders, and traders' utility functions that capture their valuations under social influence. One can extend the equilibrium theorem of Arrow and Debreu to prove the existence of equilibria in this market model with social influence.

The goal of this paper is to understand and characterize the impact of social influence on the complexity of computing and approximating market equilibria. To this end, we focus on two concrete influence models based on the traditional linear utility functions (to be defined below). We present both complexity-theoretic and algorithmic results for approximating an equilibrium in the two settings. Our complexity results show that even a bounded-degree and planar social network can significantly increase the difficulty of equilibrium computation even in markets only a constant number of goods. (Recall that a market equilibrium in a conventional Arrow-Debrau exchange market with linear utilities can be computed in polynomial time by convex programming [11, 23].) Our algorithmic results suggest that finding an approximate equilibrium in markets with hierarchical influence networks might be easier than that in markets with arbitrary neighborhood structures.

Our study of the market model with social influence is inspired by the work of Megiddo [22] on the formation of preferences and strategic analysis. In his work on game theory [22], Megiddo argued that, in some situations, players have to start analyzing the game before they have formed their preferences over the outcomes of the game. However, strategic analysis naturally depends on preferences. He presented a generalization of Nash's equilibrium theorem to resolve this dilemma, proving that equilibrium strategies exist in his more general model of non-cooperative games.

When extending his view from game theory to mathematical economics, where traders' preferences are their utilities, traders have to start analyzing the market before they have completely formed their utilities, and their utility functions depend on what their social neighbors have. The extension of Arrow-Debreu's equilibrium theorem in the market model with social influence can be viewed as an analog of Megiddo's equilibrium theorem for games.

While our primary intended contribution is to understand the impact of social influence on the complexity of market equilibria, we also intend to challenge a common belief that the equilibria of markets with a constant number of goods are easy to compute or easy to approximate, due to the success in [7] and [9]. We refer to this belief as the *myth of a constant number of goods.* By demonstrating a simple market with a constant number of goods and a bounded-degree, planar influence graph whose equilibrium is PPAD-hard to approximate, we provide a natural counterexample to this belief. This example also helps to enhance our own appreciation of the results in [7, 9].

We also made a few technical contributions in this paper. We consider two concrete influence functions based on the linear utility functions. In our linear influence model, we allow neighbors' possessions to influence the slopes of a trader's linear utility function. In our threshold influence model, we allow neighbors' possessions to influence the thresholds of a trader's additively separable and piecewise-linear utility function. For both models, we show that finding a approximate market equilibrium is PPAD-hard. In these proofs we introduce several schematic refinements to the work of Chen et al [3, 5] that computing an equilibrium in a market with additively separable, piecewise-linear and concave utilities is PPAD-hard. In particular, for the linear influence model we prove that even when there are only four goods and the influence network is a bounded-degree, planar graph, the equilibrium approximation problem is still PPAD-hard. Indeed both proofs work for the special case of Fisher's model [2, 5, 25]. We hope our PPAD-hardness constructions will help to resolve more complexity-theoretic questions concerning other exchange markets.

Algorithmically, we present a divide-&-conquer algorithm for computing an approximate equilibrium in an Arrow-Debreu market with a constant number of goods and whose influence network is hierarchical. Let $m$ denote the number of traders. We show that if a market with a hierarchical influence network has a constant number of goods and has an equilibrium in which the magnitude of the price of every good is at least $1/m^a$ for some constant $a > 0$, then a polynomially-precised-approximate market equilibrium can be found in time $m^{O(\log m)}$. Hence, although our algorithm is relatively simple, it offers a contrasting example to our complexity results on markets with arbitrary influence networks. In that case, we show that it is PPAD-hard to compute a polynomially-precised approximate equilibrium, even though the market has only four goods and its influence graph is planar with a bounded degree, and all four prices are roughly 1/4 in any of its approximate equilibrium. In other words, equilibrium approximation in markets with hierarchical influence networks might be easier than that in markets with arbitrary influence networks.

We hope our work is a step towards characterizing the impact of social influence on the complexity of computing and approximating market equilibria.

## 2  The Model and Our Main Results

We let $\mathcal{G} = \{G_1, \ldots, G_h\}$ denote a set of $h$ divisible goods and $\mathcal{T} = \{T_1, \ldots, T_m\}$ denote a set of $m$ traders. For each trader $T_k$, $k \in [m]$, we use $\mathbf{w}_k \in \mathbb{R}_+^h$ to denote her initial endowment and $\mathbf{x}_k \in \mathbb{R}_+^h$ to denote her allocation after the exchange. We always assume that the total supply of each good $G_j \in \mathcal{G}$ in the market is $\Theta(1)$, i.e., $\sum_{k \in [m]} w_{k,j} = \Theta(1)$ for all $j \in [h]$. We call the variable $x_{k,j}$ in $\mathbf{x}_k$, $j \in [h]$, the allocation variable of $T_k$ for $G_j$ or simply the $G_j$-variable of $T_k$.

Each trader $T_k$ also has a utility function $u_k$. In the classical Arrow-Debreu market model [1], $u_k$ only depends on the allocation variables $x_{k,j}$ of $T_k$ and is a function from $\mathbb{R}_+^h$ to $\mathbb{R}_+$. In this paper we consider a more general market model with *social influence*. The major difference is that the utility function $u_k$ of $T_k$ depends on not only her own allocation $\mathbf{x}_k$ but also the allocations of other traders in the market. In general, $u_k$ could be a function over all the allocation variables: $u_k : (\mathbb{R}_+^h)^m \to \mathbb{R}_+$. We then call the tuple $\mathcal{M} = (\mathcal{G}, \mathcal{T}, \mathbf{w}_k, u_k : k \in [m])$ an *Arrow-Debreu market with social influence*.

Market equilibria can be defined similarly for this model. For convenience, we follow the convention and use $\mathbf{x}_{-k}$ to denote the $m-1$ allocation vectors $(\mathbf{x}_1, \ldots, \mathbf{x}_{k-1}, \mathbf{x}_{k+1}, \ldots, \mathbf{x}_m)$.

**Definition 1** (Equilibria in Markets with Social Influence). *A market equilibrium in a market $\mathcal{M}$ with social influence is a price vector $\mathbf{p} \in \mathbb{R}_+^h$ together with allocations $\mathbf{x}_1, \ldots, \mathbf{x}_m \in \mathbb{R}_+^h$ such that*

1. *The market* clears*: for every good $G_j \in \mathcal{G}$, we have $\sum_{k \in [m]} x_{k,j} = \sum_{k \in [m]} w_{k,j}$.*

2. *Every trader gets a* budget-feasible *and* optimal *bundle: for every $T_k \in \mathcal{T}$, we have*

    - $\mathbf{x}_k \cdot \mathbf{p} \leq \mathbf{w}_k \cdot \mathbf{p}$*; and*
    - $u_k(\mathbf{x}_k, \mathbf{x}_{-k}) \geq u_k(\mathbf{x}, \mathbf{x}_{-k})$ *for all $\mathbf{x} \in \mathbb{R}_+^h$ such that $\mathbf{x} \cdot \mathbf{p} \leq \mathbf{w}_k \cdot \mathbf{p}$.*

The mathematical properties of this model and its extensions have been extensively studied in economic theory. An equilibrium always exists under mild conditions. In particular, it always exists for the two classes of utility functions (to be defined in Section 2.2) considered in the paper. We state the existence theorem in Section 2.2 and include the proof in Appendix A for completeness.

### 2.1  Social Influence Graphs

Given a market $\mathcal{M}$ with social influence, we define its social influence graph $G_{\mathcal{M}}$ as follows: The vertex set is $\mathcal{T}$, the set of traders; there is a direct edge from $T_j$ to $T_k$ if and only if the utility $u_k$ of $T_k$ depends on (or is influenced by) at least one allocation variable of $T_j$. We use $N(T_k)$ to denote the set of predecessors of $T_k$ in $G_{\mathcal{M}}$. We call $N(T_k)$ the *influencing neighbors* of trader $T_k$.

In Section 3 and 4, we study the complexity of computing an approximate market equilibrium (see the definition below) in a market with various families of influence graphs. To make this search problem more concrete, we will focus on the following two types of utility functions.

## 2.2 Linear and Threshold Influence Utility Functions

Let $\mathcal{M}$ be a market and $G_\mathcal{M}$ be its social influence graph. Consider trader $T_k \in \mathcal{T}$.

**Definition 2** (Linear Influence Functions). *We call the utility function $u_k(\mathbf{x}_k, \mathbf{x}_j : T_j \in N(T_k))$ of trader $T_k$ a* linear influence function *if it has the following form:*

$$u_k\big(\mathbf{x}_k, \mathbf{x}_j : T_j \in N(T_k)\big) = \sum_{i \in [h]} \Big(c_{k,i} + f_{k,i}\big(\mathbf{x}_j : T_j \in N(T_k)\big)\Big) x_{k,i}, \tag{1}$$

*where $c_{k,i} \geq 0$ for all $i \in [h]$ and $f_{k,i}$ is a linear form over the allocation variables $\{\mathbf{x}_j : T_j \in N(T_k)\}$ with non-negative[1] weights.*

Note that once $\{\mathbf{x}_j : T_j \in N(T_k)\}$ are fixed, the utility function $u_k$ becomes a linear function.

**Definition 3** (Threshold Influence Functions). *We call the utility function $u_k(\mathbf{x}_k, \mathbf{x}_j : T_j \in N(T_k))$ of $T_k$ a* threshold influence function *if it has the following form:*

$$u_k\big(\mathbf{x}_k, \mathbf{x}_j : T_j \in N(T_k)\big) = \sum_{i \in [h]} \Big(c_{k,i} \cdot x_{k,i} + \min\Big(0, f_{k,i}\big(\mathbf{x}_j : T_j \in N(T_k)\big) - d_{k,i} \cdot x_{k,i}\Big)\Big), \tag{2}$$

*where $c_{k,i} \geq d_{k,i} \geq 0$ for all $i \in [h]$ and $f_{k,i}$ is a linear form over the allocation variables $\{\mathbf{x}_j : T_j \in N(T_k)\}$ with non-negative weights.*

Note that once $\{\mathbf{x}_j : T_j \in N(T_k)\}$ are fixed, $u_k$ becomes an additively separable and piecewise linear function:

$$u_k(\mathbf{x}) = u_{k,1}(x_{k,1}) + \cdots + u_{k,h}(x_{k,h}).$$

For each $i \in [h]$, either $u_{k,i}$ is a linear function with slope $c_{k,i}$ (when $d_{k,i} = 0$); or $u_{k,i}$ is piecewise-linear and the slopes of the two segments are $c_{k,i}$ and $c_{k,i} - d_{k,i}$ respectively (when $d_{k,i} > 0$), with the threshold being $f_{k,i}/d_{k,i}$.

Since scaling $u_k$ does not affect the preference of $T_k$ over different bundles, we always assume that the linear and threshold influence utility functions are *normalized*: All the parameters in (1) and (2) (including $c_{k,i}$, $d_{k,i}$ as well as the weights in $f_{k,i}$) are between 0 and 1.

For these two classes of utility functions, a market equilibrium always exists under mild conditions. We include the proof of Theorem 1 below in Appendix A for completeness. The proof also implies that finding an *approximate* equilibrium (see the definition and discussion in Section 2.3) is in the class TFNP. We need the following definition.

**Definition 4** (Nonsatiation). *A function $f$ over $\mathbb{R}_+^\ell$ is said to be* nonsatiated, *if for all $\mathbf{x} \in \mathbb{R}_+^\ell$, there exists an $\mathbf{x}' \in \mathbb{R}_+^\ell$ such that $f(\mathbf{x}) < f(\mathbf{x}')$. $f$ is said to be* nonsatiated with respect to the $i^{\text{th}}$ variable *if for all $\mathbf{x} \in \mathbb{R}_+^\ell$, there exists an $\mathbf{x}' \in \mathbb{R}_+^\ell$ with $x'_j = x_j$ for all $j \neq i$ such that $f(\mathbf{x}) < f(\mathbf{x}')$.*

---

[1] Of course in general the weights can be negative but then the sufficient conditions in Theorem 1 have to be changed accordingly. In this paper we only focus on the non-negative case.

**Definition 5** (Economy Graphs). *Given a market $\mathcal{M}$ with social influence, we use $H_\mathcal{M}$ to denote the following directed graph. The vertex set is $\mathcal{T}$, the set of traders in $\mathcal{M}$. For every two traders $T_j, T_k \in \mathcal{T}$, we have an edge from $T_j$ to $T_k$ if there exists an integer $i \in [h]$ such that $w_{j,i} > 0$ and and $u_k$ is nonsatiated with respect to $x_{k,i}$. $H_\mathcal{M}$ is then called the* economy graph *of $\mathcal{M}$ [6, 21].*

Note that if the utilities in $\mathcal{M}$ are linear influence functions, then there is an edge from $T_j$ to $T_k$ iff there exists an $i \in [h]$ such that $w_{j,i} > 0$ and $c_{k,i} > 0$. Similarly, if the utilities are threshold influence functions, then there is an edge from $T_j$ to $T_k$ iff there exists an $i \in [h]$ such that $w_{j,i} > 0$ and $c_{k.i} - d_{k,i} > 0$.

**Theorem 1** (Existence). *A market $\mathcal{M}$ with linear influence (or threshold influence) functions has a market equilibrium if the following two conditions hold:*

1. *The economy graph $H_\mathcal{M}$ is strongly connected; and*

2. *For every $G_i \in \mathcal{G}$, there exists a $T_k \in \mathcal{T}$ such that $u_k$ is* nonsatiated *with respect to $x_{k,i}$.*

### 2.3 Approximation of Market Equilibria

In both our complexity-theoretic and algorithmic studies, we use the following notion of (weakly) approximate equilibria.

**Definition 6** ($\epsilon$-Approximate Equilibria). *$(\mathbf{p}, \mathbf{x}_1, \ldots, \mathbf{x}_m)$ is an $\epsilon$-approximate equilibrium if*

1. $\sum_{i \in [h]} p_i = 1$;

2. *For every trader $T_k \in \mathcal{T}$, her budget is* approximately *feasible: $\mathbf{x}_k \cdot \mathbf{p} \leq \mathbf{w}_k \cdot \mathbf{p} + \epsilon$;*

3. *For every trader $T_k \in \mathcal{T}$, her allocation $\mathbf{x}_k$ is approximately* optimal:

$$u_k(\mathbf{x}, \mathbf{x}_{-k}) \leq u_k(\mathbf{x}_k, \mathbf{x}_{-k}) + \epsilon, \quad \text{for any } \mathbf{x} \in \mathbb{R}_+^h \text{ such that } \mathbf{x} \cdot \mathbf{p} \leq \mathbf{w}_k \cdot \mathbf{p};$$

4. *The market is* approximately *cleared: For every good $G_i \in \mathcal{G}$,*

$$\left| \sum_{k \in [m]} w_{k,i} - \sum_{k \in [m]} x_{k,i} \right| \leq \epsilon.$$

**Remark 1.** *The main reason why we adopted the weaker approximation notion above, instead of using the standard one in which both conditions 2 and 3 above must hold exactly, is to guarantee the existence of a rational $\epsilon$-approximate market equilibrium for any $\epsilon > 0$ so that the search problems considered in Section 3 and 4 are well defined. It is not clear to us whether a market with* rational *linear/threshold influence functions always has a* rational *equilibrium, or even an approximate one if conditions 2 and 3 must hold exactly. With the approximation notion above, one can follow the proof of Theorem 1 to show that the following problem belongs to* TFNP:

> *Given a market $\mathcal{M}$ with linear (or threshold) influence functions, find an $\epsilon$-approximate market equilibrium with $\epsilon = 2^{-\max(h,m)}$.*

In Section 3, we prove the following PPAD-hardness results.

1. In Section 3.2, we show that the problem of finding an $m^{-16}$-approximate equilibrium in a market with linear influence utilities is PPAD-hard, even when the market has only four goods and the social influence graph is bounded-degree and planar. This contrasts with the classical linear Arrow-Debreu model, for which efficient algorithms are known [11, 23, 7, 8, 17, 15, 18, 10, 27].

2. In Section 3.3, we show that when arbitrarily many goods are allowed, finding an $m^{-11}$-approximate equilibrium in a market with threshold influence utilities is PPAD-hard, even when the influence graph is bounded-degree and planar.

Both PPAD-hardness results actually hold for the special Fisher's market model.

In Section 4, we show that if the social influence graph $G_\mathcal{M}$ of $\mathcal{M}$ has a *hierarchical* structure (see definition in Section 4.2); and the number of goods in $\mathcal{M}$ is a constant; and $\mathcal{M}$ has a market equilibrium in which $1/p_i$ is bounded above by $m^a$ for some constant $a > 0$ for all $i \in [h]$, then a $1/\text{poly}(m)$-approximate market equilibrium can be found in time $m^{O(\log(m))}$.

## 3 Hardness of Markets with Social Influence

### 3.1 Two-Player Matrix Games and Their Nash Equilibria

A two-player game is defined by a pair of payoff matrices $(\mathbf{A}, \mathbf{B})$ of its two players. Here we assume both players have $n$ choices of actions, so $\mathbf{A}$ and $\mathbf{B}$ are square matrices with $n$ rows and columns. (In this section, we will always use $n$ to denote the size of $\mathbf{A}$ and $\mathbf{B}$.) We let $\Delta^n \subset \mathbb{R}^n$ denote the set of probability distributions of $n$ dimensions.

A pair of probability distributions $(\mathbf{x}, \mathbf{y})$, $\mathbf{x}, \mathbf{y} \in \Delta^n$, is a Nash equilibrium of $(\mathbf{A}, \mathbf{B})$, if

$$\mathbf{A}_i \mathbf{y}^T < \mathbf{A}_j \mathbf{y}^T \implies x_i = 0 \quad \text{and} \quad \mathbf{x}\mathbf{B}_i < \mathbf{x}\mathbf{B}_j \implies y_i = 0, \quad \text{for all } i, j \in [n].$$

where we let $\mathbf{A}_i$ and $\mathbf{B}_i$ denote the $i$th row vector of $\mathbf{A}$ and $i$th column vector of $\mathbf{B}$, respectively.

**Definition 7** (Well-Supported Nash Equilibria). *For $\epsilon > 0$, we say $(\mathbf{x}, \mathbf{y})$ is an $\epsilon$-well-supported Nash equilibrium of $(\mathbf{A}, \mathbf{B})$, if $\mathbf{x}, \mathbf{y} \in \Delta^n$ and for all $i, j \in [n]$, we have*

$$\mathbf{A}_i \mathbf{y}^T + \epsilon < \mathbf{A}_j \mathbf{y}^T \implies x_i = 0 \quad \text{and} \quad \mathbf{x}\mathbf{B}_i + \epsilon < \mathbf{x}\mathbf{B}_j \implies y_i = 0. \tag{3}$$

**Definition 8** (Normalized and Sparse Two-Player Games). *A two-player game $(\mathbf{A}, \mathbf{B})$ is* normalized *if every entry of $\mathbf{A}$ and $\mathbf{B}$ is between $-1$ and $1$. It is* sparse *if every row and every column of $\mathbf{A}$ and $\mathbf{B}$ have at most $10$ nonzero entries.*

Let SPARSE-NASH denote the problem of finding an $n^{-6}$-well-supported Nash equilibrium in an $n \times n$ sparse and normalized two-player game, where each payoff entry is a rational number that is specified as the ratio of two integers. We will use the following hardness result:

**Theorem 2** (Sparse Two-Player Nash [4]). SPARSE-NASH *is* PPAD-complete.

## 3.2 Markets with Linear Influence Utilities

Building on the scheme introduced in [3, 5], we reduce SPARSE-NASH to the following problem:

> PLANAR-LINEAR-MARKET: The input is a market $\mathcal{M} = (\mathcal{G}, \mathcal{T}, \mathbf{w}_k, u_k)$ satisfying all the conditions of Theorem 1, in which
> 
> 1. The number of goods $|\mathcal{G}|$ is at most 4;
> 2. Every utility function $u_k$ is a (normalized) linear influence function; and
> 3. The influence graph $G_\mathcal{M}$ is a bounded-degree, planar graph.
> 
> The output is then an $m^{-16}$-approximate market equilibrium, where $m = |\mathcal{T}|$.

**Theorem 3** (Main). PLANAR-LINEAR-MARKET *is* PPAD-hard.

### 3.2.1 The Construction

Let $(\mathbf{A}, \mathbf{B})$ be an $n \times n$ normalized and sparse two-player game. We first construct from $(\mathbf{A}, \mathbf{B})$, in polynomial time, a market $\mathcal{M}$ with only *two* goods and linear influence utilities, such that every $1/n^{15}$-approximate market equilibrium of $\mathcal{M}$ gives us a $1/n^6$-well-supported Nash equilibrium of $(\mathbf{A}, \mathbf{B})$. However, its influence graph $G_\mathcal{M}$ might not be planar. In Section 3.2.3, we show how to revise the construction so that the influence graph is planar.

Let $\alpha = 1/n^3$, $\beta = 1/n^{10}$ and $\gamma = 1/n^4$. The market $\mathcal{M}$ consists of the following traders:

$$\left\{ T, X_i, Y_i, U_{i,j}, V_{i,j}, A_{i,k}, B_{i,k} : i \in [n], j \in [n], i \neq j, k \in [n-2] \right\}.$$

They have the following initial endowments: (1) the initial endowment of $T$ is $(1,1)$, one unit of each good; and (2) the initial endowment of any other trader is $(\alpha, \alpha)$. As a result, the total supply of each good is $1 + O(1/n)$.

**Remark 2.** *The market constructed here is in fact a Fisher market (with social influence).*

**Utility Function of $T$:**

The utility of $T$ only depends on her own allocation. We set the slopes of both goods to be 1.

**Utility Function of $X_i$, $i \in [n]$:**

Other than her own allocation, the utility function of $X_i$ also depends on that of $A_{i,1}$. For convenience we use $a_{i,1}$ to denote the $G_1$-allocation variable of $A_{i,1}$. Then we set the parameters of $X_i$'s utilities appropriately so that the slope of good $G_1$ is $1 + \gamma$ and the slope of good $G_2$ is $1 + a_{i,1}$.

As it will become clear later, the $G_1$-allocation variables of $X_i$, $i \in [n]$, denoted by $x_i$, will be used to encode the probability distribution of the first player in the two-player game $(\mathbf{A}, \mathbf{B})$.

**Utility Function of $Y_i$, $i \in [n]$:**

Other than her own allocation, the utility of $Y_i$ only depends on that of $B_{i,1}$. We let $b_{i,1}$ denote the $G_1$-allocation variable of $B_{i,1}$. Then we set the parameters of $Y_i$'s utilities appropriately so that the slope of good $G_1$ is $1 + \gamma$ and the slope of good $G_2$ is $1 + b_{i,1}$.

As it will become clear later, the $G_1$-allocation variables of $Y_i$, $i \in [n]$, denoted by $y_i$, will be used to encode the probability distribution of the second player in the two-player game $(\mathbf{A}, \mathbf{B})$.

**Utility Functions of $U_{i,j}$ and $V_{i,j}$, $i \neq j \in [n]$:**

The role of $U_{i,j}$ is to enforce the following Nash equilibrium constraint:

$$\mathbf{A}_i \mathbf{y}^T + \epsilon < \mathbf{A}_j \mathbf{y}^T \implies x_i = 0,$$

where, as mentioned above, we use $x_i$ to denote the $G_1$-allocation variable of $X_i$ and $y_i$ to denote the $G_1$-allocation variable of $Y_i$.

To this end, we let

$$C_{i,\ell} = \frac{\max(0, A_{i,\ell} - A_{j,\ell})}{2} \quad \text{and} \quad D_{i,\ell} = \frac{\max(0, A_{j,\ell} - A_{i,\ell})}{2}, \quad \text{for every } \ell \in [n].$$

Hence we have $0 \leq C_{i,\ell}, D_{i,\ell} \leq 1$ and $\mathbf{A}_i - \mathbf{A}_j = 2(\mathbf{C}_i - \mathbf{D}_j)$.

The utility function of $U_{i,j}$ depends on $y_\ell$, the $G_1$-allocation variable of $Y_\ell$ with

$$A_{i,\ell} - A_{j,\ell} = 2(C_{i,\ell} - D_{i,\ell}) \neq 0.$$

Since $\mathbf{A}$ is a sparse matrix, there can be at most 20 such $\ell$'s. We set the parameters of $U_{i,j}$'s utility appropriately so that the slopes of the two goods are

$$1 + \mathbf{D} \cdot \mathbf{y}^T \quad \text{and} \quad 1 + \beta + \mathbf{C} \cdot \mathbf{y}^T,$$

respectively. Similarly the role of $V_{i,j}$ is to enforce the following constraint:

$$\mathbf{x}\mathbf{B}_i + \epsilon < \mathbf{x}\mathbf{B}_j \implies y_i = 0.$$

Her utility is similar to that of $U_{i,j}$ except that it depends on $x_\ell$, the $G_1$-allocation variable of $X_\ell$ with $B_{\ell,i} - B_{\ell,j} \neq 0$. We omit the details here.

**Utility Functions of $A_{i,k}$ and $B_{i,k}$, $i \in [n], k \in [n-2]$:**

The role of the traders $A_{i,k}$, $k \in [n-2]$, is the following. Let $a_{i,j}$ denote the $G_1$-allocation variable of $A_{i,j}$, and $u_{i,j}$ denote the $G_1$-allocation variable of $U_{i,j}$. Then we need to set the utilities of $A_{i,1}$, $\ldots, A_{i,n-2}$ appropriately so that in every approximate market equilibrium, we have

$$\exists j : u_{i,j} \text{ is large} \implies a_{i,1} \text{ is large} \quad \text{and} \quad \forall j : u_{i,j} \text{ is close to } 0 \implies a_{i,1} \text{ is close to } 0,$$

while keeping the degree of the influence graph small. In particular, the utility of $A_{i,k}$ only depends on two other traders.

To this end, we set the utilities as follows. Here we take $i = 1$ as an example. The general case can be done similarly. If $k = n - 2$, then the utility of $A_{1,k}$ depends on both $U_{1,n-1}$ and $U_{1,n}$. We set the slopes of her utility function to be

$$1 + u_{1,n-1} + u_{1,n} \quad \text{and} \quad 1 + \gamma.$$

For each $k < n - 1$, the utility of $A_{i,k}$ depends on both $A_{1,k+1}$ and $U_{1,k+1}$. We set the slopes of her utility function to be

$$1 + a_{1,k+1} + u_{1,k+1} \quad \text{and} \quad 1 + \gamma.$$

The utilities of $B_{i,k}$'s, $k \in [n-2]$, are set similarly, and we omit the details here.

### 3.2.2 Correctness of the Reduction

It is easy to check that the market constructed satisfies all the conditions of Theorem 1, and the degree of its influence graph is bounded by 20 (though may not be planar yet).

Now suppose we have an $\epsilon$-approximate equilibrium where $\epsilon = 1/n^{15}$ (in which the sum of the prices $p_1 + p_2$ is equal to 1).

First we show that $p_1$ and $p_2$ must be very close to $1/2$.

**Lemma 1.** *In every $\epsilon$-approximate equilibrium, $p_1, p_2 \in [1/2 - \lambda, 1/2 + \lambda]$ with $\lambda = 1/n^{14}$.*

*Proof.* Suppose this is not the case and without loss of generality, $p_1 < 1/2 - \lambda$ and $p_2 > 1/2 + \lambda$.

First the budget of $T$ is 1 since her initial endowment is $(1, 1)$. Also it is clear that the optimal bundle for her is $(1/p_1, 0)$ with utility $1/p_1$. Let $(t_1, t_2)$ be the allocation of $T$ in the $\epsilon$-approximate equilibrium, then by definition (conditions 2 and 3) we have

$$t_1 \cdot p_1 + t_2 \cdot p_2 \leq 1 + \epsilon \quad \text{and} \quad t_1 + t_2 \geq 1/p_1 - \epsilon.$$

It then follows that

$$t_2 \leq \frac{\epsilon(1 + p_1)}{p_2 - p_1} \leq \frac{2\epsilon}{2\lambda} = \frac{1}{n}.$$

However, the total budget of all other traders in the market is

$$\Theta(n^2) \cdot (\alpha \cdot p_1 + \alpha \cdot p_2) = \Theta(1/n).$$

As a result, even if they spend all the money on $G_2$, they can consume at most

$$\frac{O(1/n)}{p_2} = O\left(\frac{1}{n}\right)$$

of $G_2$ and the total consumption of $G_2$ is $O(1/n)$. This contradicts with the assumption since the total supply of $G_2$ is $1 + \Theta(1/n)$. □

Next, we let $\mathbf{x} = (x_1, \ldots, x_n)$ denote the vector in which $x_i$ is the $G_1$-allocation variable of $X_i$; and $\mathbf{y} = (y_1, \ldots, y_n)$ denote the vector in which $y_i$ is the $G_1$-allocation variable of $Y_i$, $i \in [n]$. We show that after rounding:

$$x'_i = \begin{cases} 0 & \text{if } x_i = O(1/n^{12}) \\ x_i & \text{otherwise} \end{cases} \quad \text{and} \quad y'_i = \begin{cases} 0 & \text{if } y_i = O(1/n^{12}) \\ y_i & \text{otherwise} \end{cases}$$

and normalization:

$$x^*_i = \frac{x'_i}{\sum_{i \in [n]} x'_i} \quad \text{and} \quad y^*_i = \frac{y'_i}{\sum_{i \in [n]} y'_i},$$

the pair of distributions $(\mathbf{x}^*, \mathbf{y}^*)$ must be a $1/n^6$-well-supported Nash equilibrium of $(\mathbf{A}, \mathbf{B})$. To this end, we need the following two lemmas.

**Lemma 2.** *Let $k \in [n]$ be any index that maximizes $\mathbf{A}_k \mathbf{y}^T$, then we must have $x_k = \Omega(1/n^3)$; Let $k \in [n]$ be any index that maximizes $\mathbf{x} \mathbf{B}_k$, then we must have $y_k = \Omega(1/n^3)$.*

*Proof.* Without loss of generality, we assume $\mathbf{A}_1 \mathbf{y}^T = \max_k \mathbf{A}_k \mathbf{y}^T$.

Now we examine traders $U_{1,j}$, $j \in [2:n]$. For each $j \in [2:n]$, by the construction, the slopes of the two goods $G_1, G_2$ of $U_{1,j}$ are

$$1 + \mathbf{D}\mathbf{y}^T \quad \text{and} \quad 1 + \beta + \mathbf{C}\mathbf{y}^T,$$

respectively, where

$$(\mathbf{C} - \mathbf{D})\mathbf{y}^T = (\mathbf{A}_1 \mathbf{y}^T - \mathbf{A}_j \mathbf{y}^T)/2 \geq 0.$$

Since the budget of $U_{1,j}$ is $\alpha$, her optimal bundle is $(0, \alpha/p_2)$ with utility $\alpha(1 + \beta + \mathbf{C}\mathbf{y}^T)/p_2$. Let $(s_1, s_2)$ denote the allocation of $U_{1,j}$ in the approximate equilibrium, then by definition we have

$$s_1 \cdot p_1 + s_2 \cdot p_2 \leq \alpha + \epsilon \quad \text{and} \quad s_1(1 + \mathbf{D}\mathbf{y}^T) + s_2(1 + \beta + \mathbf{C}\mathbf{y}^T) \geq \alpha(1 + \beta + \mathbf{C}\mathbf{y}^T)/p_2 - \epsilon.$$

It then follows that $s_1 = O(\epsilon/\beta) = O(1/n^5) \ll \gamma$.

Similarly it can be shown by induction that for every trader $A_{1,j}$, $j \in [n-2]$, her $G_1$-allocation variable $a_{1,j}$ satisfies $a_{1,j} = O(\epsilon/\gamma) = O(1/n^{11})$ and in particular, $a_{1,1} = O(1/n^{11})$. With this, the slopes of trader $X_1$ becomes

$$1 + \gamma = 1 + 1/n^4 \quad \text{and} \quad 1 + O(1/n^{11})$$

for the two goods, respectively. Using the same argument, it is easy to show that

$$x_1 = \Omega\left(\frac{\alpha}{p_1}\right) = \Omega\left(\frac{1}{n^3}\right).$$

The second part for $\mathbf{y}$ can be proved similarly. $\square$

**Lemma 3.** *For all $i \neq j \in [n]$, we have*

$$\mathbf{A}_i \mathbf{y}^T + 4/n^{10} < \mathbf{A}_j \mathbf{y}^T \implies x_i = O(1/n^{12}) \quad \text{and} \quad \mathbf{x}\mathbf{B}_i + 4/n^{10} < \mathbf{x}\mathbf{B}_j \implies y_i = O(1/n^{12}).$$

*Proof.* Without loss of generality, we assume $(\mathbf{A}_1 - \mathbf{A}_j)\mathbf{y}^T < -4/n^{10}$ for some $j \neq 1$.

First of all, the slopes of trader $U_{1,j}$ are

$$1 + \mathbf{D}\mathbf{y}^T \quad \text{and} \quad 1 + 1/n^{10} + \mathbf{C}\mathbf{y}^T,$$

respectively. Since $(\mathbf{C} - \mathbf{D})\mathbf{y}^T = (\mathbf{A}_1 - \mathbf{A}_j)\mathbf{y}^T/2 < -2/n^{10}$, her optimal bundle is $(\alpha/p_1, 0)$ with utility $\alpha(1 + \mathbf{D}\mathbf{y}^T)/p_1$. It is then easy to show that in an $\epsilon$-approximate equilibrium her allocation of $G_1$ must satisfy $u_{1,j} = \Omega(1/n^3)$.

Next one can show that, by induction, all traders $A_{1,1}, \ldots, A_{1,j-1}$ like $G_1$ much better than $G_2$ and in any $\epsilon$-approximate market equilibrium, we have $a_{1,1} = \Omega(1/n^3)$. As a result, trader $X_1$ likes $G_2$ much better than $G_1$ and it can be shown that $x_1 = O(\epsilon \cdot n^3) = O(1/n^{12})$. □

By combining the two lemmas above, it is easy to show that $(\mathbf{x}^*, \mathbf{y}^*)$ is a $1/n^6$-well-supported Nash equilibrium of $(\mathbf{A}, \mathbf{B})$, proving the correctness of the reduction.

### 3.2.3 Reduction to Markets with Planar Influence Graphs

We next show that the hardness result remains to hold even if the influence graph is *planar*. Let $(\mathbf{A}, \mathbf{B})$ be a sparse two-player game and $\mathcal{M}$ be the market constructed above. We set

$$\alpha = 1/n^9, \quad \beta = 1/n^{16}, \quad \text{and} \quad \gamma = 1/n^{10}.$$

We also add two new goods to the market $\mathcal{M}$ so that there are totally four goods $G_1, G_2, G_3, G_4$. We change the initial endowments and utilities of the traders in $\mathcal{M}$ as follows:

1. The initial endowment of $T$ is now $(1, 1, 1, 1)$ (and thus, her budget is still 1);

2. The initial endowment of any other trader in $\mathcal{M}$ is $(\alpha, \alpha, \alpha, \alpha)$ (and thus, her budget is $\alpha$);

3. The utility function of $T$ now has slope 1 for all of the four goods; and the utility of any other trader remains unchanged (so they are only interested in goods $G_1$ and $G_2$).

Let $G_\mathcal{M}$ be the influence graph of $\mathcal{M}$. We first compute (in polynomial time) a planar embedding of the directed graph $G_\mathcal{M}$ so that no three directed edges intersect at the same point. Let $\mathcal{S}$ denote the set of all intersections in this planar embedding. We then add a new trader for each intersection $S \in \mathcal{S}$ and call her trader $S$. For each direct edge $T_i T_j$ in $G_\mathcal{M}$, letting $T_i S_1, S_1 S_2, \ldots, S_\ell T_j$ be the segments along $T_i T_j$ in the embedding, we add a new trader for each segment $S_k S_{k+1}$, $k \in [\ell - 1]$, and for $S_\ell T_j$. We call them trader $S_k S_{k+1}$ and trader $S_\ell T_j$, respectively, for convenience.

The total number of traders is bounded by $O(n^8)$ since there are only $O(n^2)$ traders in $\mathcal{M}$. All the new traders have the same initial endowment $(\alpha, \alpha, \alpha, \alpha)$ (and thus, all of them have budget $\alpha$). Let $\epsilon = 1/n^{32}$. Then given any $\epsilon$-approximate equilibrium of this new market (even though we have not set the utilities of the new traders yet), we can show the following lemma concerning the price vector $\mathbf{p}$ (with $\sum_{i \in [4]} p_i = 1$). The proof is similar to that of Lemma 1 so we omit it here.

**Lemma 4.** *In any $\epsilon$-approximate equilibrium, $p_1, p_2, p_3, p_4 \in [1/4 - \lambda, 1/4 + \lambda]$ where $\lambda = 1/n^{31}$.*

Next we set the utilities of the new traders appropriately so that along each directed edge $T_i T_j$ of $G_{\mathcal{M}}$ with segments $T_i S_1, S_1 S_2, \ldots, S_\ell T_j$, the $G_1$-allocation variable of $T_i$ is "almost faithfully" copied along the edge, by the $G_1$-allocation variables of traders $S_1 S_2, \ldots, S_{\ell-1} S_\ell$ and $S_\ell T_j$ finally. As a result, we are able to use the $G_1$-allocation variable of $S_\ell T_j$, instead of that of $T_i$, to influence the utility of $T_j$, by replacing the $G_1$-variable of $T_i$ in the utility function of $T_j$ with the $G_1$-variable of $S_\ell T_j$. The goal is to preserve the original reduction while making the influence graph of the new market planar. To this end, we set the utilities of the new traders as follows.

Let $S_1, S_2, S_3, S_4$ and $S$ be five intersection points in $\mathcal{S}$ such that $S_1 S, S S_2$ are on the same edge of $G_{\mathcal{M}}$ and $S_3 S, S S_4$ are on the same edge of $G_{\mathcal{M}}$. For convenience, we also use $s_{1,i}, s_{2,i}, s_{3,i}$ and $s_{4,i}$ to denote the $G_i$-variable of traders $S_1 S, S S_2, S_3 S$ and $S S_4$, respectively, and $s_i$ to denote the $G_i$-variable of $S$. We set the utility functions of $S, S_3$ and $S_4$ as follows so that in any $\epsilon$-approximate market equilibrium, $s_{2,1}$ is *very* close to $s_{1,1}$ and $s_{4,1}$ is *very* close to $s_{3,1}$.

1. The utility of $S$ depends on all traders $S_1 S, S S_2, S_3 S$ and $S S_4$. The slopes of the goods are:

$$1 + s_{1,1} + s_{2,2}, \quad 1 + s_{2,1} + s_{1,2}, \quad 1 + s_{3,1} + s_{4,2} \quad \text{and} \quad 1 + s_{4,1} + s_{3,2}.$$

2. The utility of $S S_2$ only depends on $S$. The slopes of the four goods are:

$$1 + s_1, \quad 1 + s_2, \quad 0 \quad \text{and} \quad 0.$$

3. The utility of $S S_4$ only depends on $S$. The slopes of the four goods are:

$$1 + s_3, \quad 1 + s_4, \quad 0 \quad \text{and} \quad 0.$$

Lemma 5 then follows from simple case analysis.

**Lemma 5.** *In any $\epsilon$-approximate equilibrium, we have $|s_{2,1} - s_{1,1}|, |s_{4,1} - s_{3,1}| \leq 1/n^{21}$.*

This completes the construction. With Lemma 5, one can prove new versions of Lemma 2 and Lemma 3. Using the fact that there more than $n^2$ traders in the market, Theorem 3 then follows.

### 3.3 Markets with Threshold Influence Utilities

In this section, we reduce SPARSE-NASH to the following market equilibrium problem:

> PLANAR-THRESHOLD-MARKET: The input is a market $\mathcal{M} = (\mathcal{G}, \mathcal{T}, \mathbf{w}_i, u_i)$ satisfying all conditions of Theorem 1, in which every utility is a normalized threshold influence function and the influence graph $G_{\mathcal{M}}$ is planar and has constant degree. The output is an $m^{-11}$-approximate market equilibrium, where $m = |\mathcal{T}|$.

**Theorem 4.** PLANAR-THRESHOLD-MARKET *is* PPAD-hard.

To this end, we use the polynomial-time reduction from SPARSE-NASH to *classical* Arrow-Debreu markets with additively separable and piecewise linear utility functions presented in [3].

We first briefly review the reduction, and then revise it to get a reduction from SPARSE-NASH to PLANAR-THRESHOLD-MARKET. Now let $(\mathbf{A}, \mathbf{B})$ be an $n \times n$ sparse two-player game, then the classical Arrow-Debreu market constructed in [3] has $h = 2n + 2$ goods and $m = \Theta(n^2)$ traders:

1. For every trader $T$, her utility function $u$ is additively separable: $u(\mathbf{x}) = \sum_{i \in [h]} u_i(x_i)$, where $\mathbf{x} = (x_1, \ldots, x_h)$ is the allocation of $T$. Every $u_i$ is either the zero function or a 2-piecewise-linear function:

    $$u_i(x) = a_i \cdot x, \text{ for } x \in [0, \theta_i] \quad \text{and} \quad u_i(x) = a_i \cdot \theta + b_i(x - \theta_i), \text{ for } x > \theta_i,$$

    for some $a_i \geq b_i > 0$ and $0 \leq \theta_i \leq 1/n^4$.

2. Every utility function $u$ is *sparse*: There are only constant many $i \in [h]$ such that $u_i$ is not the zero function.

3. The economy graph of the market is strongly connected.

4. The total supply of good $G_i$, for every $i \in [h]$, is $\Theta(1)$.

Moreover, given any $n^{-13}$-(strongly-)approximate market equilibrium of $\mathcal{M}$, one can construct an $n^{-6}$-well-supported Nash equilibrium of $(\mathbf{A}, \mathbf{B})$ in polynomial time. (Note that the approximation notion used in [3] is stronger, in which both conditions 2 and 3 of Definition 6 must hold exactly.)

Now using $\mathcal{M}$, we construct a new market $\mathcal{M}^*$ with threshold influence utility functions, whose influence graph is both constant-degree and planar: At the beginning $\mathcal{M}^*$ is empty. Then for every trader $T$ in $\mathcal{M}$ with initial endowment $\mathbf{w}$ and utility function $u(\mathbf{x}) = \sum_{i \in [h]} u_i(x_i)$, (we know that every $u_i$ is either the zero function or a 2-piecewise-linear function with parameters $(a_i, b_i, \theta_i)$) we let $S$ denote the set of $i \in [h]$ such that $u_i$ is not the zero function.

> Create $|S| + 1$ traders $T^*, T_i^*, i \in S$, in the new market $\mathcal{M}^*$. For each $i \in S$, the initial endowment of $T_i^*$ is $(1/n^4) \cdot \mathbf{e}_i$, where $\mathbf{e}_i$ denotes the $i$th unit vector. Let $\mathbf{x} = (x_1, \ldots, x_h)$ denote the allocation variables of $T_i^*$, then her utility function is simply $u(\mathbf{x}) = x_i$. The initial endowment of $T^*$ is set to be $\mathbf{w}$ and the utility function $u^*$ of $T^*$ depends on the allocation variables of $T_i^*$, $i \in [S]$. We set the parameters of her utility function appropriately so that $u^*$ is *exactly* $u$ (the utility of $T$ in $\mathcal{M}$) if for every $i \in [S]$, the allocation of $T_i^*$ is *exactly* $(1/n^4) \cdot \mathbf{e}_i$.

This finishes the construction of $\mathcal{M}^*$ from $(\mathbf{A}, \mathbf{B})$.

It is easy to check that the influence graph of $\mathcal{M}^*$ is constant-degree and planar. The economy graph of $\mathcal{M}^*$ is strongly connected, and the total supply of $G_i$, $i \in [h]$, is still $\Theta(1)$.

Now we sketch the proof of correctness. First of all, one can prove that in any $n^{-22}$-(weakly-)approximate equilibrium of $\mathcal{M}^*$ the allocation of any trader $T_i^*$ (as in the construction above) must be very close to $(1/n^4)\mathbf{e}_i$ and thus, the utility function of $T^*$ is very close to that of $T$ in $\mathcal{M}$. As a result, the preferences of the traders $T^*$ in $\mathcal{M}^*$ are very similar to those of traders $T$ in $\mathcal{M}$. One can then follow the proof of [3] closely to show that, given any $n^{-22}$-(weakly)-approximate equilibrium of $\mathcal{M}^*$, an $n^{-6}$-well-supported Nash equilibrium of $(\mathbf{A},\mathbf{B})$ can be computed in polynomial time. Theorem 4 then follows from the fact that $m = \Theta(n^2)$.

## 4 Algorithm for Markets with Hierarchical Influence Graphs

In this section, we consider the special case when the market $\mathcal{M}$ has only a constant number of goods and the underlying influence graph is *hierarchical*. The utility of every trader in $\mathcal{M}$ is either a linear influence function or a threshold influence function, and the market satisfies all conditions of Theorem 1. We use $h = |\mathcal{G}|$ to denote the number of goods, which is a constant, and use $m = |\mathcal{T}|$ to denote the number of traders. Then we show that

**Theorem 5.** *If $\mathcal{M}$ has an equilibrium $(\mathbf{p}, \mathbf{x}_1, \ldots, \mathbf{x}_m)$ in which for every $i \in [h]$, $1/p_i$ is bounded above by $m^a$ for some constant $a$, then a $1/\text{poly}(m)$-approximate market equilibrium of $\mathcal{M}$ can be found in time $m^{O(\log m)}$.*

Notice that even for markets with a constant-degree and planar influence graph, this problem (with the guarantee that $\mathcal{M}$ has an equilibrium in which none of the prices is negligible) is PPAD-hard, as implied by the polynomial-time reduction presented in Section 3.2 (since we know all the four prices must be $1/4$ in any equilibrium).

### 4.1 Existence of Approximate Equilibria with Discrete Prices

Let $\mathcal{M}$ be a market with (normalized) linear and threshold utility functions. For convenience, we assume the total supply of each good is exactly 1.

We assume $\mathcal{M}$ has an equilibrium $(\mathbf{p}^*, \mathbf{x}_1^*, \ldots, \mathbf{x}_m^*)$ with $p_i^* > 1/m^a$ for some constant $a > 0$. Let $\epsilon = 1/m^b$ for some constant $b > 0$. Then we can round the equilibrium $(\mathbf{p}^*, \mathbf{x}_1^*, \ldots, \mathbf{x}_m^*)$:

$$p_i = \frac{\lceil p_i^* \cdot N \rceil}{N} \quad \text{and} \quad x_{k,i}^* = \frac{\lceil x_{k,i}^* \cdot N \rceil}{N}, \quad \text{where } N = \lceil m^{2a+b+3} \rceil,$$

to get a new tuple $(\mathbf{p}, \mathbf{x}_1, \ldots, \mathbf{x}_m)$ in which every entry is a multiple of $1/N$. Using the definition, it can be checked that $(\mathbf{p}, \mathbf{x}_1, \ldots, \mathbf{x}_m)$ must be an $\epsilon$-approximate market equilibrium[2] of $\mathcal{M}$ with

$$1 \le \sum_{i \in [h]} p_i \le 2 \quad \text{and} \quad \sum_{k \in [m]} x_{k,i} \le 2, \quad \text{for all } i \in [h]. \tag{4}$$

For convenience, we say a vector is *discrete* if all of its entries are multiples of $1/N$.

As a consequence, to find an $\epsilon$-approximate equilibrium of $\mathcal{M}$, we only need to enumerate all discrete price vectors $\mathbf{p}$ that satisfy (4). For each vector $\mathbf{p}$ we check whether there exists a *discrete tuple* $(\mathbf{x}_1, \ldots, \mathbf{x}_m)$, satisfying (4), such that $(\mathbf{p}, \mathbf{x}_1, \ldots, \mathbf{x}_m)$ is an $\epsilon$-approximate equilibrium of $\mathcal{M}$.

---
[2]More exactly, we need to normalize the vector $\mathbf{p}$ so that $\sum_{i \in [h]} p_i = 1$.

Notably there are only polynomially many vectors $\mathbf{p}$ to check, when $h$ is a constant. Next we show for every $\mathbf{p}$, the checking can be done in time $m^{O(\log m)}$ for trees and hierarchical influence graphs. Theorem 5 then follows.

## 4.2 Trees and Hierarchical Influence Graphs

Let $\mathbf{p}$ be any discrete price vector that satisfies (4).

We start with the simplest case when the influence graph $G_\mathcal{M}$ is a complete binary tree. Every node $v$ in the tree is a trader, and every edge is bidirectional (and thus, the influence between two connected traders is also bidirectional). Due to the tree structure, every trader $v$ can influence (at most) three traders in the market. We use $\mathbf{w}_v$ to denote the initial endowment of $v$, with budget $b_v = \mathbf{w}_v \cdot \mathbf{p}$.

To verify whether $\mathbf{p}$ is an $\epsilon$-approximate market equilibrium price vector, we use the following top-down and divide-and-conquer algorithm Check-Tree$(\mathcal{T}, \mathbf{x}, \mathbf{y})$:

1. $\mathcal{T}$ is a complete binary tree with $r$ as its root; and
2. Both $\mathbf{x}$ and $\mathbf{y}$ are $h$-dimensional discrete vectors.

The algorithm then returns a collection of vector $\{\mathbf{x}_v : v \in \mathcal{T}\}$ such that

1. $\mathbf{x}_r = \mathbf{x}$ and $\sum_{v \in \mathcal{T}} \mathbf{x}_v = \mathbf{y}$; and
2. For every $v \in \mathcal{T}$ (including $r$), $\mathbf{x}_v$ is an $\epsilon$-approximately feasible and $\epsilon$-approximately optimal bundle for $v$ with respect to $\mathbf{p}$ and her neighbors' allocations, if such a tuple exists; and the algorithm returns 'nil' otherwise.

To verify whether $\mathbf{p}$ is an $\epsilon$-approximate equilibrium price, we only need to call the Check-Tree with $\mathcal{T}$ being the whole binary tree, and $\mathbf{x}, \mathbf{y}$ being all possible discrete vectors satisfying (4). This gives us an algorithm with time complexity $(2N)^{2h}$ times the complexity of Check-Tree.

Check-Tree runs recursively as follows:

1. The case when $\mathcal{T}$ is a single node is trivial: One simply checks whether $\mathbf{x} = \mathbf{y}$ and $\mathbf{x}$ is also approximately budget-feasible and approximately optimal for the trader.

2. Otherwise, let $r$ be the root of $\mathcal{T}$ and let $\mathcal{T}_1$ and $\mathcal{T}_2$ be the two subtrees of $r$. We then enumerate all possibilities of $\mathbf{x}_1$ for the root of $\mathcal{T}_1$, denoted by $r_1$; $\mathbf{x}_2$ for the root of $\mathcal{T}_2$, denoted by $r_2$; $\mathbf{y}_1$ for $\mathcal{T}_1$ as the total consumption of $\mathcal{T}_1$; and $\mathbf{y}_2$ as the total consumption of $\mathcal{T}_2$, such that

    (a) $\mathbf{x}_r = \mathbf{x}$ and $\mathbf{x}_r + \mathbf{y}_1 + \mathbf{y}_2 = \mathbf{y}$;
    (b) $\mathbf{x}_r = \mathbf{x}$ is approximately feasible and optimal for $r$, given $\mathbf{p}$ as the price vector and $\mathbf{x}_1$ and $\mathbf{x}_2$ as the allocations of $r_1$ and $r_2$, respectively.

For any combination of $(\mathbf{x}_1, \mathbf{x}_2, \mathbf{y}_1, \mathbf{y}_2)$ that satisfies the conditions above, we modify the utility of $r_1$ and $r_2$ by replacing the allocation variables of $r$, in the utility functions of $r_1$ and $r_2$, with $\mathbf{x}_r$. Then we recursively call Check-Tree$(\mathcal{T}_1, \mathbf{x}_1, \mathbf{y}_1)$ and Check-Tree$(\mathcal{T}_2, \mathbf{x}_2, \mathbf{y}_2)$. If neither call returns 'nil', we concatenate the outputs with $\mathbf{x}_r$ and output.

If we use $\ell$, the depth of the tree, to measure the time complexity TIME$(h)$ of Check-Tree, then

$$\text{TIME}(\ell) = \text{TIME}(\ell - 1) \cdot N^{O(1)}.$$

Thus, TIME$(\ell) = N^{O(\ell)}$ and hence the running time of Check-Tree measured using $m$, the number of traders in the market, is $m^{O(\log m)}$.

Clearly this divide-and-conquer approach can be applied to any market with a constant-degree tree influence graph. More generally, it can be applied to the following family of graphs which we call *hierarchical* graphs:

**Definition 9** (Hierarchical Graphs). *We call $G$ a $k$-hierarchical graph, if there is a tree $T$ such that we can label every vertex $v$ in $G$ with a node $w$ in $T$:*

1. *For every $w$ in $T$ that is not a leaf, the number of vertices in $G$ labeled with $w$ is between 1 and $k$; (If $w$ is a leaf, then this number can be arbitrarily large.)*

2. *There is an edge from $v_1$ to $v_2$ and from $v_2$ to $v_1$ if $v_1$ and $v_2$ has the same lable and the label is not a leaf in $T$;*

3. *There is an edge from $v_1$ to $v_2$ and from $v_2$ to $v_1$ if the labels of $v_1, v_2$ are neighbors in $T$.*

When $k$ and the degree of the underlying tree are constant, the divide-and-conquer approach also yields an algorithm for finding a $1/\text{poly}(m)$-approximate market equilibrium with time complexity $m^{O(\log m)}$, when the market has an equilibrium in which the price of each good is bounded below by $1/n^a$, for some constant $a$.

# References

[1] K. J. Arrow and G. Debreu. Existence of an equilibrium for a competitive economy. *Econometrica*, 22:265–290, 1954.

[2] W.C. Brainard and H.E. Scarf. How to compute equilibrium prices in 1891. Cowles Foundation Discussion Papers 1272, Cowles Foundation, Yale University, 2000.

[3] X. Chen, D. Dai, Y. Du, and S.-H. Teng. Settling the complexity of Arrow-Debreu equilibria in markets with additively separable utilities. In *Proceedings of the 50th Annual IEEE Symposium on Foundations of Computer Science*, pages 273–282, 2009.

[4] X. Chen, X. Deng, and S.-H. Teng. Settling the complexity of computing two-player nash equilibria. *J. ACM*, 56(3):1–57, 2009.

[5] X. Chen and S.-H. Teng. Spending is not easier than trading: On the computational equivalence of Fisher and Arrow-Debreu equilibria. In *Proceedings of the 20th International Symposium on Algorithms and Computation*, pages 647–656, 2009.

[6] B. Codenotti, B. McCune, S. Penumatcha, and K. Varadarajan. Market equilibrium for CES exchange economies: Existence, multiplicity, and computation. In *Proceedings of the 25th Conference on Foundations of Software Technology and Theoretical Computer Science*, pages 505–516, 2005.

[7] X. Deng, C.H. Papadimitriou, and S. Safra. On the complexity of price equilibria. *Journal of Computer and System Sciences*, 67(2):311–324, 2003.

[8] N. R. Devanur, C. H. Papadimitriou, A. Saberi, and V. V. Vazirani. Market equilibria via a primal-dual-type algorithm. In *Proceedings of the 43th Annual IEEE Symposium on Foundations of Computer Science*, pages 389–395, 2002.

[9] N. R. Devanur and R. Kannan. Market equilibria in polynomial time for fixed number of goods or agents. In *Proceedings of the 49th annual IEEE Symposium on Foundations of Computer Science*, pages 45–53, 2008.

[10] N. R. Devanur and V. V. Vazirani. An improved approximation scheme for computing Arrow-Debreu prices for the linear case. In *Proceedings of the 23rd Conference on Foundations of Software Technology and Theoretical Computer Science*, pages 149–155, 2003.

[11] E. Eisenberg and D. Gale. Consensus of subjective probabilities: The pari-mutuel method. *Annals Of Mathematical Statistics*, 30(1):165–168, 1959.

[12] I.V. Evstigneev and M. Taksar. Stochastic equilibria on graphs, ii. *Journal of Mathematical Economics*, 24:383–406, 1995.

[13] I.V. Evstigneev and M. Taksar. Stochastic economies with locally interacting agents. Working papers, Santa Fe Institute, 2001.

[14] H. Föllmer. Random economies with many interacting agents. *Journal of Mathematical Economics*, 1:51–62, 1974.

[15] R. Garg and S. Kapoor. Auction algorithms for market equilibrium. In *Proceedings of the 36th annual ACM symposium on Theory of computing*, pages 511–518, 2004.

[16] U. Horsta and J.A. Scheinkman. Equilibria in systems of social interactions. *Journal of Economic Theory*, 130(1):44–77, 2006.

[17] K. Jain. A polynomial time algorithm for computing the arrow-debreu market equilibrium for linear utilities. In *Proceedings of the 45th Annual IEEE Symposium on Foundations of Computer Science*, pages 286–294, 2004.

[18] K. Jain, M. Mahdian, and A. Saberi. Approximating market equilibria. In *Proceedings of the 6th International Workshop on Approximation Algorithms*, pages 98–108, 2003.

[19] S. Kakutani. A generalization of Brouwer's fixed point theorem. *Duke Mathematical Journal*, 8:457–459, 1941.

[20] A. Kirman. Demand theory and general equilibrium: From explanation to introspection, a journey down the wrong road. Economics working papers, Institute for Advanced Study, School of Social Science, 2006.

[21] R. R. Maxfield. General equilibrium and the theory of directed graphs. *Journal of Mathematical Economics*, 27(1):23–51, 1997.

[22] N. Megiddo. Formation of preferences and strategic analysis: Can they be de-coupled? *Presented at IMGTA 2001 – XIV Italian Meeting on Game Theory and Applications*, 2003.

[23] E. I. Nenakov and M. E. Primak. One algorithm for finding solutions of the Arrow-Debreu model. *Kibernetica*, 3:127–128, 1983.

[24] H. Scarf. *The Computation of Economic Equilibria*. Yale University Press, 1973.

[25] V. V. Vazirani and M. Yannakakis. Market equilibrium under separable, piecewise-linear, concave utilities. In *Proceedings of the 1st Symposium on Innovations in Computer Science*, 2010.

[26] L. Walras. *Elements of Pure Economics, or the Theory of Social Wealth*. 1954 English Translation, 1874.

[27] Y. Ye. A path to the Arrow-Debreu competitive market equilibrium. *Mathematical Programming*, 111:315–348, 2008.

## A   Existence of Equilibria in Markets with Social Influence

In this section, we prove Theorem 1. We will use the following fixed point theorem [19]:

**Theorem 6** (Kakutani). *Let $S$ be a non-empty, compact and convex subset of $\mathbb{R}^n$. Let $\phi : S \to 2^S$ be an upper semicontinuous correspondence such that $\phi(x)$ is non-empty, closed and convex for all $x \in S$. Then $\phi$ has a fixed point $x^* \in S$ such that $x^* \in \phi(x^*)$.*

To prove Theorem 1 using Kakutani's fixed point theorem, we follow the strategy used by Scarf [24] and recently by Vazirani and Yannakakis [25]. Given a market, we construct a correspondence $\phi$ over the product space of prices and allocations. Then we show that $\phi$ satisfies all the conditions of Kakutani's fixed point theorem. Moreover, every fixed point of $\phi$ must be an equilibrium of the market. Theorem 1 then follows.

## A.1 The Correspondence $\phi$

Recall that the total supply of each good $G_i$ in the market is 1, and all parameters ($w_{k,i}$, $c_{k,i}$, $d_{k,i}$ as well as the weights in $f_{k,i}$) are between 0 and 1.

We let $L$ be a large enough integer so that $2^L \geq 4mh^2$ and for all $k \in [m]$ and $i \in [h]$,

$$\text{either } c_{k,i} - d_{k,i} = 0 \text{ or } c_{k,i} - d_{k,i} \geq 2^{-L}.$$

The domain of the correspondence $\phi$ is $[0, 1.1]^{hm} \times P$, where $P$ is the space of prices $\mathbf{p}$:

$$S = \left\{ \mathbf{p} \in \mathbb{R}_+^h \,\Big|\, \sum_{i \in [h]} p_i = 1 \text{ and } p_i \geq c \text{ for all } i \in [h] \right\},$$

where $c = 1/(m \cdot 2^{3mL})$. (As will become clear later, we require all prices to be positive ($\geq c$) to make sure that $\phi$ is semicontinuous.) $\phi$ maps $(\mathbf{x}_1, \ldots, \mathbf{x}_m, \mathbf{p})$, where $\mathbf{x}_k \in [0, 1.1]^h$ for all $k \in [m]$ and $\mathbf{p} \in P$, to the set of all tuples $(\mathbf{x}_1^*, \ldots, \mathbf{x}_m^*, \mathbf{p}^*)$ that satisfy the following conditions:

1. Let $\mathbf{x} = \sum_{k \in [m]} \mathbf{x}_k$, then $\mathbf{p}^* \in P$ and $\mathbf{x} \cdot \mathbf{p}^* \geq \mathbf{x} \cdot \mathbf{p}'$ for all $\mathbf{p}' \in P$; and

2. For each $k \in [m]$, $\mathbf{x}_k^* \in [0, 1.1]^h$ is one of the optimal budget-feasible bundles for $T_k$:

   (a) $\mathbf{x}_k^* \cdot \mathbf{p} \leq \mathbf{w}_k \cdot \mathbf{p}$; and

   (b) $u_k(\mathbf{x}_k^*, \mathbf{x}_{-k}) \geq u_k(\mathbf{x}_k', \mathbf{x}_{-k})$ for all $\mathbf{x}_k' \in [0, 1.1]^h$ such that $\mathbf{x}_k' \cdot \mathbf{p} \leq \mathbf{w}_k \cdot \mathbf{p}$.

This completes the definition of $\phi$.

One can verify that, if the utility functions $u_k$ satisfy all the conditions of Theorem 1, then $\phi$ satisfies all the conditions of Kakutani's fixed point theorem, and has at least one fixed point.

## A.2 Every Fixed Point is a Market Equilibrium

Let $(\mathbf{x}_1, \ldots, \mathbf{x}_m, \mathbf{p})$ be a fixed point: $(\mathbf{x}_1, \ldots, \mathbf{x}_m, \mathbf{p}) \in \phi(\mathbf{x}_1, \ldots, \mathbf{x}_m, \mathbf{p})$. Also let $\mathbf{x} = \sum_{k \in [m]} \mathbf{x}_k$.

We divide the proof into several lemmas. First we show that $x_i \leq 1$ for all $i \in [h]$.

**Lemma 6.** *If $(\mathbf{x}_1, \ldots, \mathbf{x}_m, \mathbf{p})$ is a fixed point of $\phi$, then $x_i < 1.1$ for all $i \in [h]$.*

*Proof.* If $x_i = 1.1$ for some $i \in [h]$, then by the optimality of $\mathbf{p}$ we must have $x_j = 1.1$ for all $j \in [h]$ such that $p_j > c$. As a result, we have

$$\sum_{k \in [m]} \mathbf{x}_k \cdot \mathbf{p} = \mathbf{x} \cdot \mathbf{p} = \sum_{j: p_j = c} x_j \cdot p_j + \sum_{j: p_j > c} x_j \cdot p_j \geq 1.1 \sum_{j: p_j > c} p_j.$$

However, $\sum_{j: p_j > c} p_j$ is at least $1 - hc$ and we have

$$1.1 \cdot (1 - hc) > 1 = \sum_{k \in [m]} \mathbf{w}_k \cdot \mathbf{p},$$

which contradicts with the assumption that every $\mathbf{x}_k$ is budget-feasible. $\square$

Then we show that $x_i > c$ for all $i \in [h]$. We need the following lemma:

**Lemma 7.** *Let $(\mathbf{x}_1, \ldots, \mathbf{x}_m, \mathbf{p})$ be any fixed point of $\phi$. Let $T_k \in \mathcal{T}$ be a trader with budget $w = \mathbf{w}_k \cdot \mathbf{p} > 0$ and her function $u_k$ is nonsatiated with respected to $G_i \in \mathcal{G}$. Then we must have*

$$p_i \geq \frac{w}{2^{2L}}.$$

*Proof.* Assume for contradiction that $p_i < w/2^{2L}$. By the assumption, we have $c_{k,i} \geq 2^{-L}$ (and $c_{k,i} - d_{k,i} \geq 2^{-L}$, for threshold influence functions), and the bang-per-buck of $G_i$ is at least

$$\frac{2^{-L}}{w/2^{2L}} = \frac{2^L}{w}.$$

On the other hand, the slope of any good $G_j$ in the market is

$$c_{k,j} + f_{k,j} \leq 1 + mh.$$

Now let $S \subseteq [h]$ be the set of index $j$ such that

$$p_j > \frac{w(1 + mh)}{2^L}.$$

For every $j \in S$, the bang-per-buck of $G_j$ is at most

$$\frac{1 + mh}{p_j} < \frac{2^L}{w},$$

and is strictly smaller than that of $G_i$. Therefore, by the optimality of $\mathbf{x}_k$, we have $x_{k,j} = 0$ for all $j \in S$ unless $x_i = 1.1$. However, by Lemma 6 we have $x_i < 1.1$ and thus, $x_{k,j} = 0$ for all $j \in S$. We then get a contradiction since the total cost of the bundle is

$$\sum_{j \notin S} x_{k,j} \cdot p_j < 1.1 \sum_{j \notin S} p_j \leq 1.1 h \cdot \frac{w(1 + mh)}{2^L} \leq \frac{4mh^2 \cdot w}{2^L} \leq w,$$

and thus, $\mathbf{x}_k$ is not optimal. $\square$

**Corollary 1.** *If $(\mathbf{x}_1, \ldots, \mathbf{x}_m, \mathbf{p})$ is a fixed point of $\phi$, then $p_i > c$ for all $i \in [h]$.*

*Proof.* Assume for contradiction that $p_1 = c$. Then by condition 3 of Theorem 1, we assume that the utility function $u_k$ of $T_k$ is nonsatiated with respect to $G_1$.

As the total budget of all the $m$ traders is 1, there must be a trader $T_{k'} \in \mathcal{T}$ whose budget is $\mathbf{w}_{k'} \cdot \mathbf{p} \geq 1/m$. Moreover, since the economy graph $H_\mathcal{M}$ is strongly connected, there is a sequence of at most $m$ traders: $T_{k'} = T_{i_0}, T_{i_1}, \ldots, T_{i,\ell-1}, T_{i,\ell} = T_k$ such that there is an edge from $T_{i_{s+1}}$ to $T_{i_s}$ for all $s : 0 \leq s \leq \ell - 1$.

We then prove the following bound on the budget of $T_{i_s}$ by induction on $s$:

$$\text{the budget of } T_{i_s} \geq \frac{1}{m \cdot 2^{3sL}}, \quad \text{for all } s : 0 \leq s \leq \ell. \tag{5}$$

The base when $s = 0$ is trivial. Assume (5) is true for $s \geq 0$. Because there is an edge from $T_{i_{s+1}}$ to $T_{i_s}$, there exists a good $G_j$ such that $w_{i_{s+1},j} > 0$ (and thus, $> 2^{-L}$ by assumption) and $u_{i_s}$ is nonsatiated with respect to $G_j$. Since the budget of $T_{i_s}$ is at least $1/(m \cdot 2^{3sL})$, by Lemma 7

$$p_j \geq \frac{1}{m \cdot 2^{3sL} \cdot 2^{2L}}.$$

As a result, the budget of $T_{i_{s+1}}$ is at least

$$w_{i_{s+1},j} \cdot p_j \geq \frac{1}{m \cdot 2^{3(s+1)L}}.$$

This finishes the induction.

As a result, the budget of $T_k = T_{i,\ell}$ is at least

$$\frac{1}{m \cdot 2^{3\ell L}} > \frac{1}{m \cdot 2^{3(m-1)L}},$$

contradicting with $p_1 = c$ and Lemma 7. $\square$

Now we have $p_i > c$ for all $i \in [h]$. Since $\mathbf{p}$ maximizes $\mathbf{x} \cdot \mathbf{p}$, we must have

$$x_1 = x_2 = \ldots = x_n.$$

To finish the proof, we show that they must all equal to 1. To this end, if $x_1 > 1$ then we have

$$\sum_{k \in [m]} \mathbf{x}_k \cdot \mathbf{p} = \mathbf{x} \cdot \mathbf{p} > \sum_{i \in [h]} p_i = 1 = \sum_{k \in [m]} \mathbf{w}_k \cdot \mathbf{p},$$

which contradicts the assumption that every $\mathbf{x}_k$ is budget-feasible. And if $x_1 < 1$, then

$$\mathbf{x} \cdot \mathbf{p} = x_1 \sum_{i \in [h]} p_i < 1 = \sum_{k \in [m]} \mathbf{w}_k \cdot \mathbf{p}.$$

As a result, one of the traders $T_k$ did not exhaust her budget. By the concavity and nonsatiation of $u_k$, this contradicts with the optimality of $\mathbf{x}_k$.